\documentstyle[12pt]{article}

\catcode`\@=11
\long\def\@makefntext#1{
\protect\noindent \hbox to 3.2pt {\hskip-.9pt
$^{{\ninerm\@thefnmark}}$\hfil}#1\hfill}                

\def\@makefnmark{\hbox to 0pt{$^{\@thefnmark}$\hss}}  

\def\ps@myheadings{\let\@mkboth\@gobbletwo
\def\@oddhead{\hbox{}
\rightmark\hfil\ninerm\thepage}
\def\@oddfoot{}\def\@evenhead{\ninerm\thepage\hfil
\leftmark\hbox{}}\def\@evenfoot{}
\def\sectionmark##1{}\def\subsectionmark##1{}}

\renewcommand{\thefootnote}{\fnsymbol{footnote}}

\newcounter{sectionc}
\newcounter{subsectionc}
\newcounter{subsubsectionc}
\renewcommand{\section}[1] {\vspace*{0.6cm}\addtocounter{sectionc}{1}
\setcounter{subsectionc}{0}\setcounter{subsubsectionc}{0}\noindent
        {\normalsize\bf\thesectionc. #1}\par\vspace*{0.4cm}}
\renewcommand{\subsection}[1]
{\vspace*{0.6cm}\addtocounter{subsectionc}{1}
        \setcounter{subsubsectionc}{0}\noindent
        {\normalsize\it\thesectionc.\thesubsectionc.
#1}\par\vspace*{0.4cm}}
\renewcommand{\subsubsection}[1]
{\vspace*{0.6cm}\addtocounter{subsubsectionc}{1}
        \noindent
{\normalsize\rm\thesectionc.\thesubsectionc.\thesubsubsectionc.
        #1}\par\vspace*{0.4cm}}

\newcounter{appendixc}
\newcounter{subappendixc}[appendixc]
\newcounter{subsubappendixc}[subappendixc]

\renewcommand{\appendix}[1] {\vspace*{0.6cm}
        \refstepcounter{appendixc}
        \setcounter{figure}{0}
        \setcounter{table}{0}
        \setcounter{equation}{0}
        \renewcommand{\thefigure}{\Alph{appendixc}.\arabic{figure}}
        \renewcommand{\thetable}{\Alph{appendixc}.\arabic{table}}
        \renewcommand{\theappendixc}{\Alph{appendixc}}
        \renewcommand{\theequation}{\Alph{appendixc}.\arabic{equation}}
        \noindent{\bf Appendix \theappendixc #1}\par\vspace*{0.4cm}}

\def\abstracts#1{{

\centering{\begin{minipage}{12.2truecm}%
\footnotesize\baselineskip=12pt\noindent
        \centerline{\footnotesize ABSTRACT}\vspace*{0.3cm}
        \parindent=0pt #1
        \end{minipage}}\par}}

\renewenvironment{thebibliography}[1]
        {\begin{list}{\arabic{enumi}.}
        {\usecounter{enumi}\setlength{\parsep}{0pt}
\setlength{\leftmargin 1.25cm}{\rightmargin 0pt}
         \setlength{\itemsep}{0pt} \settowidth
        {\labelwidth}{#1.}\sloppy}}{\end{list}}

\newcounter{itemlistc}
\newcounter{romanlistc}
\newcounter{alphlistc}
\newcounter{arabiclistc}

\newcommand{\fcaption}[1]{
        \refstepcounter{figure}
        \setbox\@tempboxa = \hbox{\footnotesize Fig.~\thefigure. #1}
        \ifdim \wd\@tempboxa > 6in
           {\begin{center}
        \parbox{6in}{\footnotesize\baselineskip=12pt Fig.~\thefigure.
#1}
            \end{center}}
        \else
             {\begin{center}
             {\footnotesize Fig.~\thefigure. #1}
              \end{center}}
        \fi}

\newcommand{\tcaption}[1]{
        \refstepcounter{table}
        \setbox\@tempboxa = \hbox{\footnotesize Table~\thetable. #1}
        \ifdim \wd\@tempboxa > 6in
           {\begin{center}
        \parbox{6in}{\footnotesize\baselineskip=12pt Table~\thetable.
#1}
            \end{center}}
        \else
             {\begin{center}
             {\footnotesize Table~\thetable. #1}
              \end{center}}
        \fi}

\def\@citex[#1]#2{\if@filesw\immediate\write\@auxout
        {\string\citation{#2}}\fi
\def\@citea{}\@cite{\@for\@citeb:=#2\do
        {\@citea\def\@citea{,}\@ifundefined
        {b@\@citeb}{{\bf ?}\@warning
        {Citation `\@citeb' on page \thepage \space undefined}}
        {\csname b@\@citeb\endcsname}}}{#1}}

\newif\if@cghi
\def\cite{\@cghitrue\@ifnextchar [{\@tempswatrue
        \@citex}{\@tempswafalse\@citex[]}}
\def\citelow{\@cghifalse\@ifnextchar [{\@tempswatrue
        \@citex}{\@tempswafalse\@citex[]}}
\def\@cite#1#2{{$\null^{#1}$\if@tempswa\typeout
        {IJCGA warning: optional citation argument
        ignored: `#2'} \fi}}

 1
 1
 1

\font\ninerm=cmr9

\def\lsim{\mathrel{\mathop{\kern 0pt <}\limits_{\displaystyle\sim}}}
\def\gsim{\mathrel{\mathop{\kern 0pt >}\limits_{\displaystyle\sim}}}
\begin{document}

\centerline{\normalsize\bf POWER CORRECTIONS FROM SHORT
DISTANCES\footnote
{Expanded version of a talk presented at the EPS HEP97 Conference
(Jerusalem, August 1997).}}
\baselineskip=22pt

\vspace{0.6cm}

\centerline{\footnotesize G. GRUNBERG\footnote{On leave of absence from
Centre de Physique Th\'eorique de l'Ecole
 Polytechnique, 91128 Palaiseau Cedex, France.} }

\baselineskip=13pt
\centerline{\footnotesize\it Theory Division, CERN, CH-1211 Geneva 23,
Switzerland}

\vspace{0.6cm}
\centerline{\footnotesize  (grunberg@pth.polytechnique.fr)}

\vspace{0.9cm}
\abstracts{\normalsize
It is argued that power contributions of short distance origin
naturally arise in the infrared finite coupling approach. A
phenomenology of $1/Q^2$ power corrections is sketched.}
\vspace{10cm}
CERN-TH/97-340
\hspace{9cm}
November 1997

\vspace{0.5cm}
\newpage
\pagestyle{plain}
\normalsize\baselineskip=15pt
\setcounter{footnote}{0}
\renewcommand{\thefootnote}{\alph{footnote}}
\section{Introduction} Power-behaved contributions to hard processes
not amenable to operator product expansion  (OPE) have been derived
\cite{Rev} in recent years through various techniques (renormalons,
finite gluon mass, dispersive approach), which all share  the
assumption that these contributions are of essentially  {\em  infrared}
(IR)  origin. In this talk, I point out that the IR finite coupling
approach \cite{DW,DMW} naturally suggests the existence of additionnal
non-standard contributions of {\em  ultraviolet} (UV) origin, hence not
related  to renormalons (but which may be connected \cite{G,AZ} to the
removal of the Landau pole from the perturbative coupling).

\section{Power corrections and IR regular coupling}
Consider the contribution to an Euclidean (quark dominated) observable
arising
from dressed virtual
single gluon  exchange, which takes the generic form (after subtraction
of the Born term):
\begin{equation}D(Q^2) =\int_{0}^\infty{dk^2\over k^2}\
\bar{\alpha}_s(k^2)\
\varphi\left(k^2\over
Q^2\right)\end{equation}
The ``physical'' coupling $\bar{\alpha}_s(k^2)$ is assumed to be IR
regular, and thus
must differ from the perturbative coupling  $\bar{\alpha}_s^{PT}(k^2)$
( which is assumed to contain a Landau pole) by a non-perturbative
piece $\delta\bar{\alpha}_s(k^2)$:
\begin{equation}\bar{\alpha}_s=\bar{\alpha}_s^{PT}+
\delta\bar{\alpha}_s\end{equation}

To determine the various types of
power contributions, it is appropriate \cite{DW,N} to disentangle long
from short distances ``a la SVZ'' with an IR cutoff  $\Lambda_I$:
\begin{equation}D(Q^2) =\int_{0}^{\Lambda_I^2}{dk^2\over k^2}\
\bar{\alpha}_s(k^2)\ \varphi\left(k^2\over Q^2\right)
+ \int_{\Lambda_I^2}^\infty{dk^2\over k^2}\ \bar{\alpha}_s^{PT}(k^2)\
\varphi\left(k^2\over Q^2\right)
+ \int_{\Lambda_I^2}^\infty{dk^2\over k^2}\ \delta\bar{\alpha}_s(k^2)\
\varphi\left(k^2\over
Q^2\right)\end{equation}
The first integral yields, for large $Q^2$, ``long distance ''
power contributions  which correspond to the standard OPE
``condensates''. If the Feynman diagram kernel $\varphi
\left(k^2\over Q^2\right)$ is ${\cal O}\left((k^2/Q^2)^n\right)$ at
small
$k^2$, this piece contributes an  ${\cal
O}\left((\Lambda^2/Q^2)^n\right)$
term from a dimension $n$ condensate, with the normalization given by a
low energy average of the IR regular coupling $\bar{\alpha}_s$. The
integral over the perturbative coupling in the short distance part
represents a form
of ``regularized perturbation theory '' (choosing the IR cut-off
$\Lambda_I$ above the  Landau pole). The last integral in
eq.(3) usually yields (unless $\delta\bar{\alpha}_s(k^2)$ is
exponentially supressed) new power
contributions at large $Q^2$ of short distance origin , unrelated to
the OPE. Assume for instance a power law decrease:
\begin{equation}\delta\bar{\alpha}_s(k^2)\simeq c\left({\Lambda^2\over
k^2}\right)^p\end{equation}
The short distance integral will then contribute a piece:
\begin{equation}\int_{Q^2}^\infty{dk^2\over k^2}\
\delta\bar{\alpha}_s(k^2)\
\varphi\left(k^2
\over Q^2\right)\simeq  A\
c\left({\Lambda^2\over Q^2}\right)^p\end{equation}
where $A\equiv\int_{Q^2}^\infty{dk^2\over k^2}\ \left({Q^2\over
k^2}\right)^p\ \varphi
\left(k^2\over Q^2\right)$ is a number. If one assumes moreover that
$p<n$, the lower integration limit in eq.(5) and in A can actually be
set to zero \cite{B}, and one gets a {\em parametrically leading}
${\cal O}\left((\Lambda^2/Q^2)^p\right)$  power contribution of UV
origin, unrelated to the OPE.

\section {Power contributions to the running coupling }
A power law decrease of $\delta\bar{\alpha}_s(k^2)$ is a  natural
expectation for a coupling which is assumed to be defined  at the
non-perturbative level, and could eventually be derived from the OPE
itself as the following QED analogy shows. In QED, the coupling
$\bar{\alpha}_s(k^2)$ should be identified, in the present dressed
single gluon exchange context, to the Gell-Mann-Low effective charge
$\bar{\alpha}$, related to the photon vacuum polarisation $\Pi (k^2)$
by:
\begin{equation}\bar{\alpha}(k^2)= {\alpha\over 1+ \alpha\
\Pi(k^2/\mu^2,\alpha)}\end{equation}
One  expects  $\Pi(k^2)$, hence $\bar{\alpha}(k^2)$,  to receive power
contributions from the OPE. Of course, this cannot happen in QED
itself, which is an IR trivial theory, but might occur in the ``large
$\beta_0$'', $N_f=-\infty$ limit of QCD . Instead of $\Pi(k^2)$, it is
convenient to introduce the related (properly normalized)
renormalisation group invariant ``Adler function'' (with the Born term
removed):
\begin{equation}R(k^2)= {1\over \beta_0}\left({d\Pi\over d\log
k^2}-{d\Pi\over d\log k^2}|\alpha=0\right)\end{equation}
which contributes the higher order terms in the renormalisation group
equation:
\begin{equation}{d\bar{\alpha}_s\over d\ln k^2} = -\beta_0
(\bar{\alpha}_s)^2\left(1+  R\right)\end{equation}
where $\beta_0$ is (minus) the one  loop beta function coefficient.
Consider now the  $N_f=-\infty$ limit in QCD. Then $R(k^2)$ is expected
to be purely non-perturbative, since in this limit the perturbative
part of $\bar{\alpha}_s$ is just the one-loop coupling
$\bar{\alpha}_s^{PT}(k^2)=1/\beta_0\ln(k^2/\Lambda^2)$.
Indeed, OPE-renormalons type arguments suggest the general structure
\cite{G1} at large $k^2$:
\begin{equation}R(k^2)=\sum_{p=1}^\infty \left(a_p \log{k^2\over
\Lambda^2}+ b_p\right) \left(\ {\Lambda^2\over
k^2}\right)^p\end{equation}
where the log enhanced power corrections reflect the presence of double
IR renormalons poles \cite{B1}. Eq.(8) with $R$ as in eq.(9) can  be
easily integrated to give:
\begin{equation}\bar{\alpha}_s(k^2)=\bar{\alpha}_s^{PT}(k^2)
+{\Lambda^2\over
k^2}\left[a_1\ \bar{\alpha}_s^{PT}(k^2) + \beta_0\ (a_1+
b_1)\left(\bar{\alpha}_s^{PT}(k^2)\right)^2 \right]+{\cal
O}\left((\Lambda^4/k^4)\right)\end{equation}
One actually expects : $a_1=b_1=0$ (corresponding to the absence of
$d=2$ gauge invariant  operator ), and $a_2=0$ (corresponding to the
gluon condensate which yields only  a single renormalon pole).
It is amusing to note that keeping only the $p=2$ (gluon condensate)
contribution in eq.(9) with $a_2=0$, eq.(8) yields :
\begin{equation}\bar{\alpha}_s(k^2)=
{1\over\beta_0\left(\ln{k^2\over\Lambda^2}+
{b_2\over 2}{\Lambda^4\over k^4}\right)}\end{equation}
which coincides with a  previously suggested ansatz \cite{Gr}  based on
different arguments.

\section {$1/Q^2$ power corrections }
The previous QED - inspired model (with $a_1=b_1=0$ ) remains of
academic interest, since it is clear that the short distance power
corrections induced in QCD observables by the OPE-generated corrections
in $\bar{\alpha}_s$ are then parametrically consistent with those
expected from the OPE, and are actually probably numerically small
compared to those originating directly from the long distance piece in
eq.(3) (although it is still an interesting question whether such short
distance contributions will not mismatch the expected OPE result for
the coefficient functions). The situation changes if one assumes
\cite{G,AZ} the existence of  $1/k^2$ power corrections of non-OPE
origin in $\bar{\alpha}_s$. Evidence for such corrections has been
found in a lattice calculation \cite{M}  of the gluon condensate, and
physical arguments have been given \cite{AZ,AZ1} for their actual
occurence. For instance, consider the case where $n=2$ in the low
energy behavior of the kernel $\varphi\left(k^2\over Q^2\right)$, i.e.
where the leading OPE contribution has dimension 4  (the gluon
condensate). Then, setting  $p=1$ in eq.(4), the  parametrically
leading power contribution will be  a  $1/Q^2$ correction of short
distance origin given by the right-hand side eq.(5) with:
\begin{equation}A\equiv\int_0^\infty{dk^2\over k^2}\ {Q^2\over k^2}\
\varphi\left(k^2\over Q^2\right)\end{equation}
Assuming further the physical coupling $\bar{\alpha}_s$ is {\em
universal} \cite{DW,DMW}, so is the {\em non-perturbative} parameter
$c$ in eq.(4). Then the process dependance of the strength of the
$1/Q^2$ correction is entirely contained in the {\em computable}
parameter $A$. In particular, it is interesting to check \cite{G2}
whether $A$ in the pseudoscalar channel is substantially larger then in
the vector channel, which could help resolve \cite{YZ} a long standing
QCD sum rule puzzle \cite{NSVZ}, in addition to provide further
evidence for $1/Q^2$ corrections. Note that the proposed mechanism is
different from the (in essence purely perturbative) one based on UV
renormalons \cite{YZ,AZ}. The latter yields \cite{PR} an enhancement
factor of $18$ already in the single renormalon chain approximation
(consistent with the present dressed single gluon exchange picture),
but is subject to unknown arbitrarily large corrections from multiple
renormalons chains, at the difference of the present argument.

\vspace{2cm}
{\bf Acknowledgments}
I am grateful to M. Beneke, Yu.L. Dokshitzer, G. Marchesini and V.I.
Zakharov for useful discussions.

\vspace{2cm}


\end{document}